\documentclass[aps,pra,twocolumn,showpacs,amsmath,amssymb,preprintnumbers,superscriptaddress,10pt,longbibliography]{revtex4-1}

\usepackage{graphicx}
\graphicspath{{figures/}}
\usepackage{SIunits}
\usepackage{hyphenat}
\usepackage[bookmarks=false]{hyperref}
\usepackage{color}
\usepackage[capitalise]{cleveref}
\crefname{section}{Sec.}{Secs.}
\Crefname{section}{Section}{Sections}
\usepackage{multirow}
\usepackage{makecell}
\usepackage{subcaption}
\usepackage{MnSymbol}

\definecolor{pink}{RGB}{255,0,255}

\definecolor{red}{rgb}{1,0,0}

\definecolor{green}{RGB}{0,212,17}

\definecolor{orange}{RGB}{255,70,0}

\definecolor{blue}{RGB}{0,0,255}

\begin{document}

\title{Protecting fiber-optic quantum key distribution sources against light-injection attacks}

\author{Anastasiya~Ponosova}
\email{nastya-aleksi@mail.ru}
\affiliation{Russian Quantum Center, Skolkovo, Moscow 121205, Russia}
\affiliation{NTI Center for Quantum Communications, National University of Science and Technology MISiS, Moscow 119049, Russia}

\author{Daria~Ruzhitskaya}
\affiliation{Russian Quantum Center, Skolkovo, Moscow 121205, Russia}
\affiliation{NTI Center for Quantum Communications, National University of Science and Technology MISiS, Moscow 119049, Russia}

\author{Poompong~Chaiwongkhot}
\affiliation{Institute for Quantum Computing, University of Waterloo, Waterloo, ON, N2L~3G1 Canada}
\affiliation{Department of Physics and Astronomy, University of Waterloo, Waterloo, ON, N2L~3G1 Canada}
\affiliation{Department of Physics, Faculty of Science, Mahidol University, Bangkok, 10400 Thailand}
\affiliation{Quantum technology foundation (Thailand), Bangkok, 10110 Thailand}

\author{Vladimir~Egorov}
\affiliation{\mbox{Leading research center for Quantum internet, ITMO University, Birzhevaya line 14, 199034 St.~Petersburg, Russia}}
\affiliation{SMARTS-Quanttelecom LLC, 6 liniya V.O. 59, 199178 St.~Petersburg, Russia}

\author{Vadim~Makarov}
\affiliation{Russian Quantum Center, Skolkovo, Moscow 121205, Russia}
\affiliation{\mbox{Shanghai Branch, National Laboratory for Physical Sciences at Microscale and CAS Center for Excellence in} \mbox{Quantum Information, University of Science and Technology of China, Shanghai 201315, People's Republic of China}}
\affiliation{NTI Center for Quantum Communications, National University of Science and Technology MISiS, Moscow 119049, Russia}

\author{Anqi~Huang}
\email{angelhuang.hn@gmail.com}
\affiliation{Institute for Quantum Information \& State Key Laboratory of High Performance Computing, College of Computer Science and Technology, National University of Defense Technology, Changsha 410073, China}

\date{\today}

\begin{abstract}
A well-protected and characterised source in a quantum key distribution system is needed for its security. Unfortunately, the source is vulnerable to light-injection attacks, such as Trojan-horse, laser-seeding, and laser-damage attacks, in which an eavesdropper actively injects bright light to hack the source unit. The hacking laser could be a high-power one that can modify properties of components via the laser-damage attack and also further help the Trojan-horse and other light-injection attacks. Here we propose a countermeasure against the light-injection attacks, consisting of an additional sacrificial component placed at the exit of the source. This component should either withstand high-power incoming light while attenuating it to a safe level that cannot modify the rest of the source, or get destroyed into a permanent high-attenuation state that breaks up the line. We demonstrate experimentally that off-the-shelf fiber-optic isolators and circulators have these desired properties, at least under attack by a continuous-wave high-power laser.
\end{abstract}

\maketitle

\section{Introduction}
\label{sec:intro}

Quantum key distribution (QKD) allows to securely establish a secret key between two remote parties, usually called Alice and Bob~\cite{bennett1984,ekert1991}. Its informational-theoretical security is based on quantum physics, instead of any computational complexity~\cite{gisin2002,scarani2009,lo2014,xu2020}. This makes QKD, in principle, unhackable even by a super-powerful quantum computer. Thus, QKD is a promising candidate for quantum-safe cryptography in the era of quantum computing that is approaching with currently feasible quantum supremacy~\cite{arute2019}. However, in practice, it is a long journey to achieve an unhackable QKD system due to imperfect devices in real life~\cite{makarov2006,qi2007,lamas-linares2007,lydersen2010a,lydersen2010b,xu2010,li2011a,wiechers2011,lydersen2011c,lydersen2011b,gerhardt2011,sun2011,jain2011,bugge2014,sajeed2015a,huang2016,makarov2016,huang2018,qian2018,huang2019,huang2020,sun2022,chaiwongkhot2022,huang2022,gao2022}. The imperfections in realistic QKD systems can be exploited by an adversary equipped with current technology to learn the secret information~\cite{lydersen2010a,gerhardt2011}.

The quantum hacking discloses the practical security performance of QKD systems, which then stimulates the community to enhance the security hardness of QKD implementation. For example, a decade ago, various loopholes were discovered at the receiver side that works on detecting quantum states received from a quantum channel~\cite{makarov2006,lydersen2010a,lydersen2010b,wiechers2011,lydersen2011c}. To defeat the attacks on the quantum-state detection, measurement-device-independent QKD (MDI QKD)~\cite{lo2012} and twin-field QKD (TF QKD)~\cite{lucamarini2018,wang2022} were proposed, in which there were no security assumptions about the quantum-state measurement. Therefore, these protocols can defeat all attacks on measurement unit. In addition, MDI QKD and TF QKD schemes with well-protected senders that prepare characterised quantum states are believed to be practically secure, eliminating the threat of quantum hacking~\cite{bennett2017}. Unfortunately, quantum hackers are ingenious---it has been shown that they can learn or even manipulate the characteristics of components in the source unit by light-injection attacks, like Trojan-horse attack~\cite{gisin2006,jain2014}, laser-seeding attack~\cite{sun2015,huang2019,pang2020}, laser-damage attack~\cite{makarov2016,huang2020}, and power-meter attack~\cite{sajeed2015}. Since the modified characteristics are often unpredictable, it is difficult to build a security model that counters these active attacks. Consequently, these attacks may be the effective tools in Eve's suitcase to crack the security of MDI QKD and TF QKD systems.

A fiber-optic isolator or circulator, which is often placed as the last component in the source unit~\cite{mo2005,huang2016a,wang2016,dixon2017,xia2019,wei2019,liu2019}, is believed to protect a fiber-based QKD system from the adversary's injecting light through a quantum channel. For example, Ref.~\onlinecite{lucamarini2015} thoroughly analyses the necessary amount of isolation as countermeasure against the Trojan-horse attack and upper-bounds the remaining information leakage. Then the security can be restored by a privacy amplification. This countermeasure is also being standardised by the European Telecommunications Standards Institute (ETSI)~\cite{ETSIQKD0010}. From this point of view, protecting the source unit by isolation components seems to be a promising solution, achieving a practically secure source, especially for MDI QKD and TF QKD. Nevertheless, the actual amount of isolation may be affected by unknown attacks on the isolating component~\cite{huang2020}. Guaranteeing the practical security of the QKD system under such realistic situation is still challenging. 

Here we show that an additional sacrificial isolation component placed at the exit of the source that is not accounted into the security model can be an effective countermeasure against the light-injection attacks. We experimentally demonstrate that when the adversary illuminates isolators and circulators with a high-power continuous-wave (c.w.)\ laser, $6.4$--$42.4~\deci\bel$ residual isolation remains, although the high-power laser temporarily or permanently decreases their isolation values by $15.2$--$34.5~\deci\bel$. Since the isolation components under the high-power attack are still able to provide the significant amount of isolation, they protect other optical components behind them in the QKD source unit from the modification by the laser-damage attack. However, since this additional isolation component, the last in the QKD source, might be affected by the eavesdropper, it should not be counted into the effective isolation needed to prevent the light-injection attacks. That is, the required isolation as countermeasure against the light-injection attacks should be calculated starting from the component after our sacrificial isolation component.

The article is structured as follows. In~\cref{sec:exp_meth} we describe the experimental setup and methodology to test the fiber-optic isolators and circulators. Measurement results are presented in \cref{sec:results}. We discuss the effects of this attack and application of this countermeasure in \cref{sec:countermeasure} and conclude in \cref{sec:conclusion}.

\section{Experimental methodology}
\label{sec:exp_meth}

\subsection{Experimental setup for testing isolators}
\label{sub:setup-isolator}

\begin{figure}
  \includegraphics{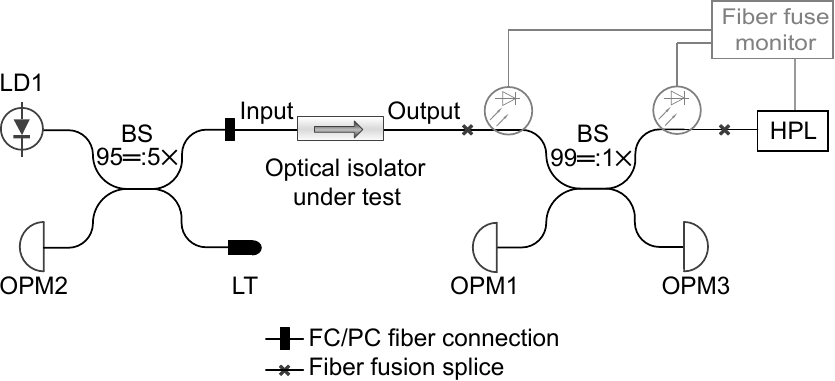}
  \caption{Experimental setup for testing isolators. LD, laser diode; OPM, optical power meter; LT, light trap; HPL, high-power laser. The coupling ratio of the beam splitter~(BS) denoted as 95$\Relbar$:5$\bigtimes$ means $95\%$ of light passes to the port horizontally opposite in the graphical symbol of the BS, while $5\%$ of light is coupled across to the other port.}
  \label{fig:setup}
\end{figure}

Our experimental setup simulates a hacking scenario in which Eve hacks the system from the quantum channel to the source unit. \Cref{fig:setup} illustrates the measurement configuration used for testing fiber-optic isolators. The samples under test are illuminated by a high-power laser~(HPL), consisting of a c.w.\ $1550~\nano\meter$ seed laser diode (QPhotonics QFBGLD-1550-100) followed by an erbium-ytterbium-doped fiber amplifier (QGLex custom-made unit)~\cite{huang2020}. The laser is transmitted through a single-mode fiber to mimic the attack via the quantum channel. As we focus on the effect of optical power on the tested sample, the polarization of laser is not characterized. Laser output power can be varied from $0.16~\watt$ to $6.7~\watt$ at the isolator under test. During the experiment, the illumination power is set by the interface of a control software according to a calibration curve made before the experiment. The optical power meter 1 (OPM1; Grandway FHP2B04), connected through the 1\% arm of a beam splitter~(BS), monitors the power emitted by the high-power laser in real time. The laser light transmitted through the samples in the backward direction is continuously monitored by OPM2 (Thorlabs PM200 with S154C sensor). The isolation is determined by comparing the power measured by OPM2 with the laser power launched into the sample, taking into account the 95:5 coupling ratio of the BS. 

A fiber-pigtailed $1550~\nano\meter$ laser diode (LD1, Gooch and Housego AA1406) with $10.5$-$\milli\watt$ optical power is used to measure the insertion loss of the isolator under test. The transmitted power is measured after 99:1 BS using OPM3 (Thorlabs PM200 with S155C sensor). The insertion loss is then determined by comparing the power measured by OPM3 with the input one, taking into account the additional $20~\deci\bel$ attenuation from the 99:1 BS. Our setup is equipped with a fiber fuse monitor, which shutdowns the high-power laser automatically in case the fiber fuse is detected, preventing an extensive damage of equipment~\cite{huang2020}. Fortunately, the fiber fuse has not occurred during the tests reported in this article. Moreover, a temperature map of the samples is measured by a thermal imaging camera (Fluke TiS45), which is placed over the samples and saves thermal images every $3~\second$ during each experiment. It is notable that during the testing on ISO PM1 as an initial trial, there is no thermal images recording yet.

\subsection{Experimental setup for testing circulators}
\label{sub:setup-circulator}

To determine the testing setup for fiber-optic circulators, we shall first discuss two configurations that a three-port circulator can have in the QKD system. In the first scenario, the circulator is employed to direct Alice's optical pulses~\cite{lucio-martinez2009,tang2014,wang2016,xia2019,liu2019}. That is, the optical pulses first pass from port 1 to port 2. Then the pluses are reflected back to port 2 and transmitted to port 3 as the output of the QKD sender. Thus, the isolation values between each port pair matter to the security of the QKD system. In the second scenario, the circulator is used to monitor the injected light~\cite{mo2005, wang2014}. If the injected light is detected by a monitor connected at port 3, Alice and Bob may interrupt their QKD session without secret key leakage. However, it has been shown that the laser-damage attack might decrease the sensitivity of the monitor~\cite{makarov2016} and the high-speed optical pulses might bypass the alarming mechanism of the monitor~\cite{sajeed2015}. In this case, the success of the light-injection attack will highly rely on the monitor’s properties and signal processing, instead of the isolation provided by the circulator, which is out of the scope of the present study.

\begin{figure}
  \includegraphics{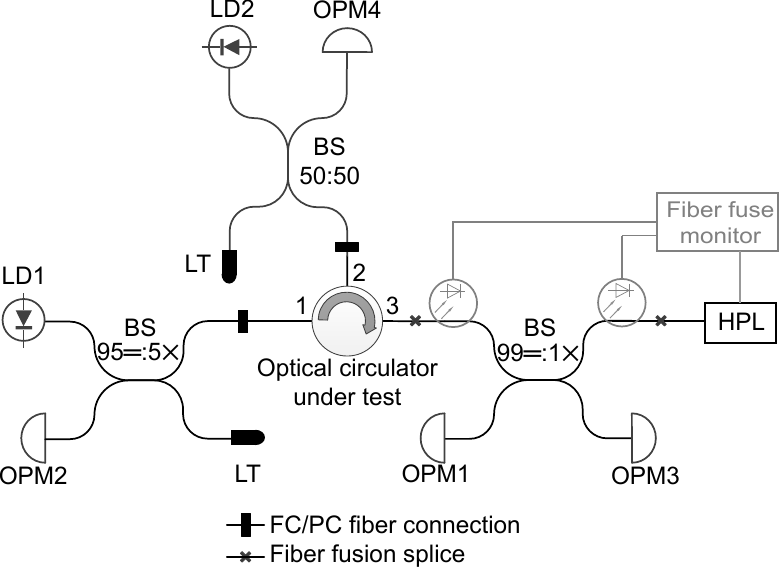}
  \caption{Experimental setup for testing circulators. LD, laser diode; OPM, optical power meter; LT, light trap; HPL, high-power laser.}
  \label{fig:CIRC_setup}
\end{figure}

As discussed above, in this study we focus on testing the isolation characterization of circulator configured in the first scenario, while testing the whole configuration in this scenario will be the future work. The experimental setup of testing the circulator is shown in~\cref{fig:CIRC_setup}. The measurement settings at ports~1 and 3 are the same as described hereinabove for isolator testing in \cref{sub:setup-isolator}. In addition to that, a laser diode (LD2, Gooch and Housego AA1406) and an optical power meter (OPM4, Thorlabs PM200 with S154C sensor) are placed at port~2 via a 50:50 BS. LD1, LD2, and the HPL are used one at a time to prevent measurement errors caused by reflected light. Isolation and insertion loss are estimated for each pair of circulator's ports via a procedure similar to that described in \cref{sub:setup-isolator}.

\subsection{Test procedure}
\label{sub:procedure}

Before starting the test on the optical isolators and circulators, we experimentally verified that up to $6.7~\watt$ none of the components in the setup, excluding the optical isolators and circulators, change their characteristics during the test. Especially, the splitting ratios of BS are not seen notably change. Thus, the only changes observed in the following test are in the isolators and circulators under test.

We define a successfully ``hacked" isolation component as one having a temporal or permanent isolation decrease without losing light transmission capability in the forward direction, within our measurement accuracy of about $1~\deci\bel$. We also notice when the insertion loss increases permanently. Such an increase would lead (with a threshold that depends on the particular QKD system) to the secret key failing to be generated. This means the eavesdropper would not be able to learn any secret information.

The test procedure is the following for each component under test. Firstly, the initial isolation and insertion loss are measured in the experimental setup before illumination by HPL. Then each sample is exposed to a constant power level starting from $0.16~\watt$ for at least $60~\second$ (except for the sample ISO~PM~1, which is exposed for at least $10~\second$ as the initial test). The exposure period may be increased up to $900~\second$ during the testing if necessary. During the illumination, the isolation of the isolator under test is monitored. For circulators, the isolation values from port 3 to port 1 and from port 3 to port 2 are measured during the illumination. If isolation reduction is detected, the laser power is kept constant until the isolation value becomes stable. After each round of illumination, the HPL is turned off, and we measured the insertion loss of the sample under test again. For isolators, LD1 is turned on, and the insertion loss is measured by OPM3. For circulators, LD1 and LD2 are turned on alternately, and insertion loss from port~1 to port~2 and from port~2 to port~3 is measured by OPM4 and OPM3. In addition, LD2 are also used to measure the isolation from port~2 to port~1 with assistance of OPM2. The temporary changes in isolation and insertion loss are recorded during the measurement.

We repeat the testing procedure above with laser power of the HPL incremented by~$100$--$500~\milli\watt$. The testing stops if an irreversible damage to the sample is incurred. For some samples, the testing stops before the sample is fully damaged. This is because we would like to measure the permanent decrease in isolation, while the sample is still operational.

\section{Results}
\label{sec:results}

\subsection{Test results for fiber-optic isolators}
\label{sub:results-isolator}

\begin{table*}
	\caption{Testing results of isolators. All measurements are at $1550~\nano\meter$.}
	\label{tab:ISO_result}
	\begin{tabular}[t]{@{\extracolsep{1.8ex}}l@{}c@{}c@{}c@{}c@{}c@{}c@{}}
		\hline\hline
		\multirow{3}{*}{Sample}
		& \multirow{3}{*}{\makecell{Specified minimum\\ isolation~($\deci\bel$)}}
		& \multicolumn{2}{c}{Initial} 
		& \multirow{3}{*}{\makecell{Minimum \\ isolation ($\deci\bel$)}}
		& \multirow{3}{*}{\makecell{Maximum decrease\\ of isolation ($\deci\bel$)}}
		& \multirow{3}{*}{\makecell{Irreversible\\ damage at}}   \\
		\cline{3-4}
		& & \makecell{Insertion\\ loss ($\deci\bel$)} &  \makecell{Isolation ($\deci\bel$)}  \\
		\hline
		ISO PM 1	& 46	& 0.66	& 53.7	& 21.8 @ $6.7~\watt$, $360~\second$		& 31.9	& $6.7~\watt$, $900~\second$	\\
		ISO PM 2	& 28	& 0.50	& 37.0	& 17.2 @ $3.37~\watt$, $820~\second$	& 19.8	& was not tested							\\
		ISO 3-1		& 46	& 0.45	& 58.1	& 37.1 @ $3.3~\watt$, $260~\second$		& 21.0	& was not tested							\\
		ISO 3-2		& 46	& 0.55	& 62.1	& 27.6 @ $3.4~\watt$, $800~\second$		& 34.5	& $3.8~\watt$, $90~\second$		\\
		ISO 4			& 55	& 0.52	& 57.6	& 42.4 @ $2.2~\watt$, $200~\second$		& 15.2	& was not tested							\\
		\hline\hline
	\end{tabular}
	\label{tab:all}
\end{table*}

We have tested four models of fiber-optic isolators used in real QKD systems: one sample of models~1,~2 and 4 (ISO~PM~1, ISO~PM~2, and ISO~4) and two samples of isolator model~3 (ISO~3-1 and ISO~3-2). All the isolators have a similar design and operation principle except that ISO~PM~1 and ISO~PM~2 are polarization-dependent, while the other two models are polarization-insensitive. According to their specifications, all the tested isolators should operate correctly at a maximum c.w.\ power of $500~\milli\watt$, except for ISO~PM~2, whose maximum operating power is $300~\milli\watt$. The operating temperature range of all the samples is $-5$ to $+70~\celsius$. Owing to our confidentiality agreements with the QKD system manufacturers, we cannot publicly disclose the part numbers of the components tested in this study. They are ordinary commercial off-the-shelf products.

\smallskip 

\begin{figure}
  \includegraphics{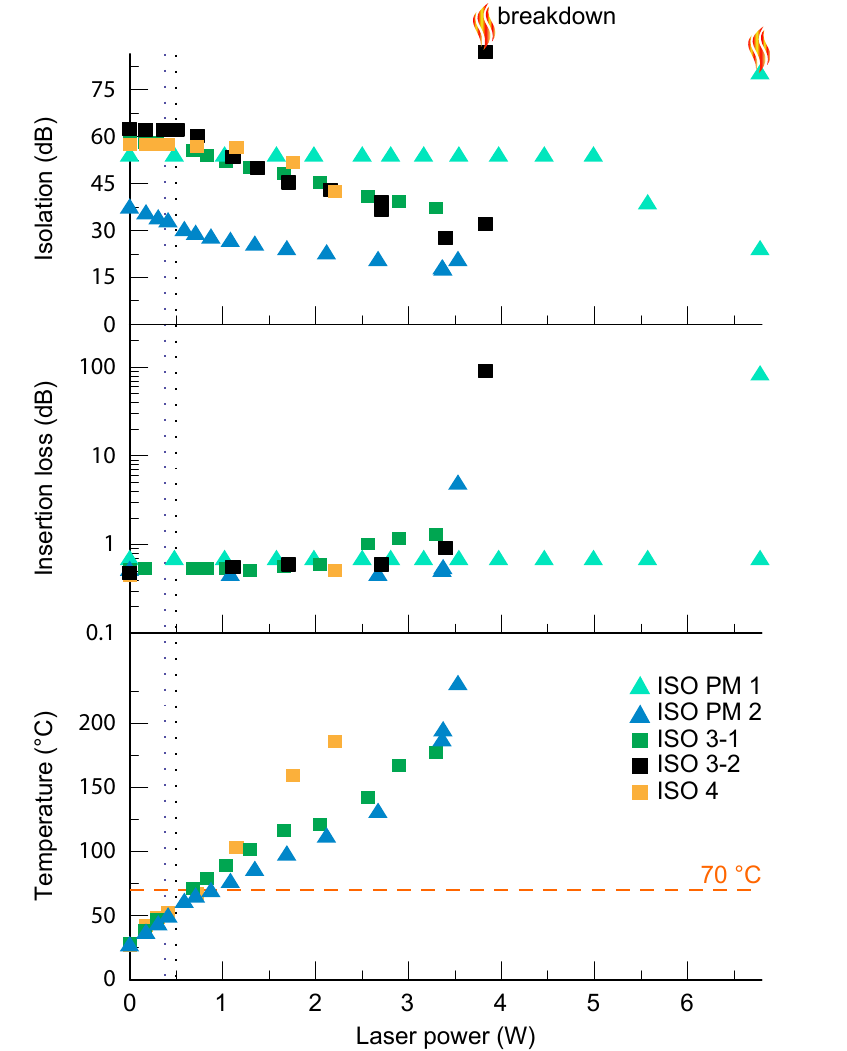}
  \caption{Isolators' parameters under testing. The points represents the minimum isolation value, maximum insertion loss value, and highest surface temperature achieved at each applied power of HPL. The temperature was only measured for three samples. The leftmost vertical dotted line is the maximum specified operating power of $300~\milli\watt$ for ISO~PM~2. The rightmost vertical dotted line is the maximum specified operating power of $500~\milli\watt$ for the other samples.}
  \label{fig:ISO_data}
\end{figure}

A summary of the laser-damage results is presented in \cref{tab:ISO_result}. The tested samples are vulnerable to the high-power injection laser, exhibiting the temporary reduction of isolation by $15.2$--$34.5$~$\deci\bel$ at a certain illumination power (see ``Maximum decrease of isolation'' in \cref{tab:ISO_result}). As a result, $17.2$--$42.4$~$\deci\bel$ isolation remains before samples become inoperable~(see ``Minimum isolation''), which is less than every sample's specified minimum isolation value. In addition, ISO~PM~1 and ISO~3-2 are destroyed at $6.7$ and $3.8~\watt$ injected laser power applied for $900$ and $90~\second$ respectively. Detailed results of the testing are given in~\cref{fig:ISO_data}.

\smallskip 

The characteristics of ISO~PM~1 differ significantly from the other samples because of the shorter laser exposure time at the beginning of its test, which is not enough to observe significant changes in isolation. However, when the exposure period is lasting longer with optical power higher than $5~\watt$, the decrease in isolation is illustrated.

\smallskip 

As can be seen from the topmost plot in \cref{fig:ISO_data}, the isolation reduction under high-power laser is observed for all the samples. It does not happen until the applied laser power exceeds the maximum operating power specified by the manufacturer, except for ISO~PM~2, for which isolation reduction from its maximum value by $3.4~\deci\bel$ was observed in the operating power range. However, even for this sample, the measured isolation conforms to the specification when the illumination laser power is in the operating range (specified minimum isolation of ISO~PM~2 is $28~\deci\bel$, see \cref{tab:all}). The ``breakdown'' points in \cref{fig:ISO_data} indicate that ISO~PM~1 and ISO~3-2 are fully damaged at the laser power of $6.7~\watt$ and $3.8~\watt$---they exhibit extremely large insertion loss and isolation. For the other samples, we stopped the laser exposure before completely destroying them, observing a permanent decrease in isolation by $3.9~\deci\bel$ for ISO~PM~2 and temporary decrease in isolation for ISO~3-1 and ISO~4.

\smallskip 

Interestingly, before being destroyed, the isolators keep operating in the forward direction (see their insertion loss values in the middle plot in \cref{fig:ISO_data}) while their isolation values are reduced. The insertion loss varies slightly by $0.5$--$1.1~\deci\bel$, which leads to the loss of only 22\% forward transmitted power at most. Once the irreversible damage happens for ISO~PM~1 and ISO~3-2, their insertion loss is larger than $80~\deci\bel$.

\begin{figure*}
	\includegraphics{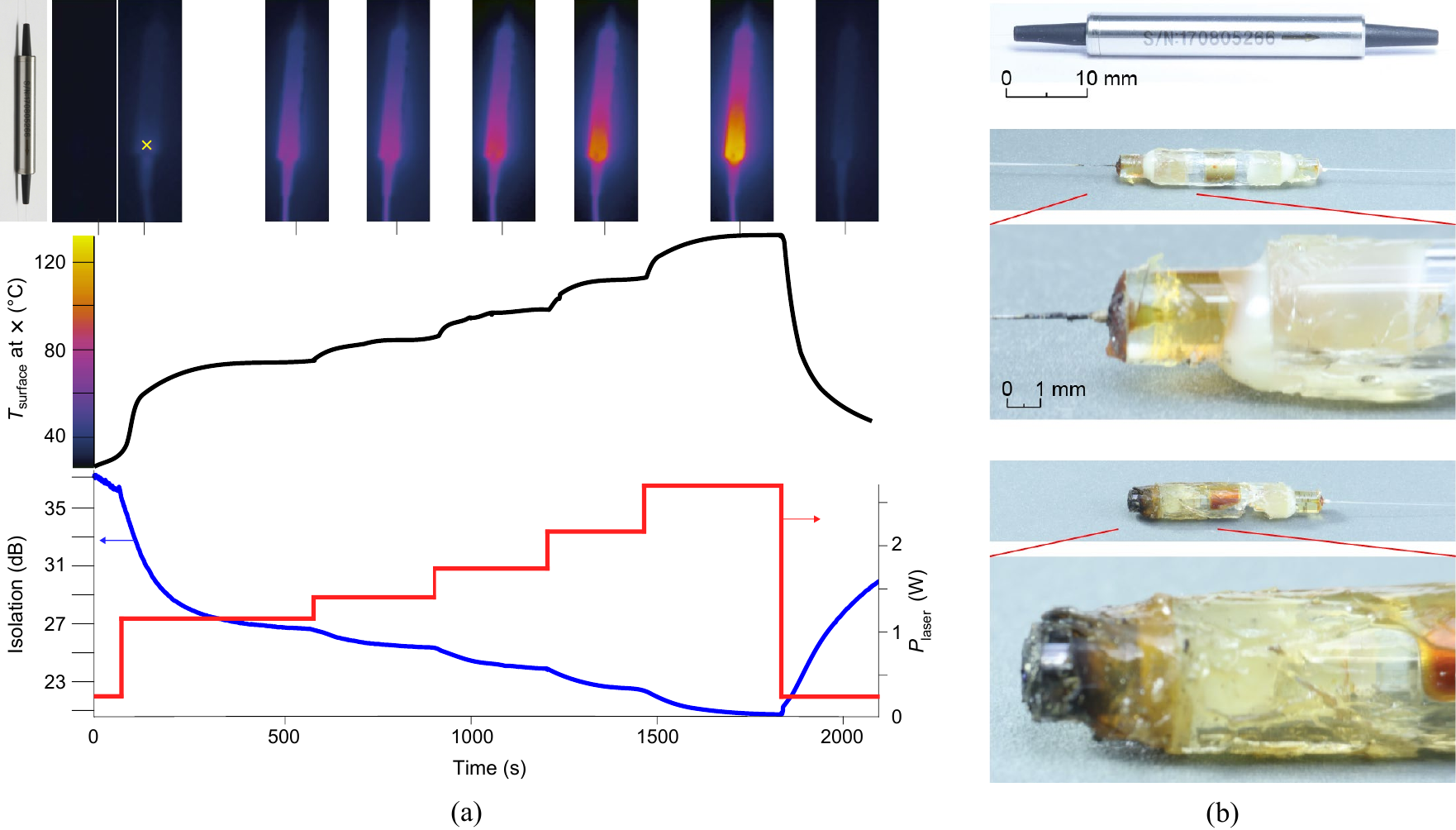}
	\caption{Analysis of isolator response to high-power laser exposure. (a)~Isolation and temperature profile of ISO~PM~2 under stepwise-increasing laser power. The temperature is measured at the hottest surface point, marked $\times$ on thermal camera images. (b)~Photograph, top to bottom: ISO~3-1 before testing, decapsulated ISO~3-1 showing its internal design (partial damage after illumination by $3.3~\watt$ is visible), decapsulated ISO~3-2 showing damage after illumination by $3.8~\watt$ laser power.}
\label{fig:ISO_trend}
\end{figure*}

\smallskip 

The sample's surface temperature (see the bottommost plot in \cref{fig:ISO_data}) rises with the illumination power. It seems to be related to the isolation value. The isolation of the polarization-insensitive samples ISO~3-1 and ISO~4 begins dropping when their measured temperature exceeds the maximum specified operating temperature of $+70~\celsius$.

\bigskip 

In order to understand the mechanism of isolation decrease and isolators' damage, we analyze the thermal images and disassemble the tested samples as shown in~\cref{fig:ISO_trend}. \Cref{fig:ISO_trend}(a) illustrates the surface temperature maps, the temperature curve, and the isolation curve of ISO~PM~2 in one experiment. The thermal profile of the isolator under a high-power laser shows that the sample is heated inhomogeneously across its surface. Specifically, the tested sample is heated at the side opposite to the input port where the high-power laser is applied, which is also observed in all the other tested samples. This is because the injected high-power laser emission is rejected to be coupled from the isolator to the optical fiber~\cite{berent2013}, and next, the rejected light is absorbed inside the package to cause this local heating.

\bigskip 

Moreover, after applying laser power higher than the sample's specified maximum operating value ($300~\milli\watt$), the amount of isolation drops rapidly with the power. After cooling, the isolation reverts close to the initial value. \Cref{fig:ISO_trend}(b) shows the external and internal design of the tested samples, ISO 3-1 and ISO 3-2. After the disassembly of the sample, we found a destroyed blackened side of the optical assembly, which matches the point of the highest surface temperature marked in the thermal images. Thus, we infer that high temperature causes this destruction.

To further verify the cause of isolation change, we theoretically simulate the working model of an isolator with details given in~\cref{sec:theory-isolation-vs-temp}. There, we have calculated temperature dependence of Verdet constant and isolation changes for a single-stage polarization-dependent isolator. The analysis shows that the polarization rotation angle depends on temperature. As a result, when the temperature becomes high, the light injected in the backward direction is not fully reflected by the isolator's polarizer but is partially transmitted. Thus, the amount of isolation is reduced under high temperature. These modeling results correlate well with the experimental data of ISO~PM~2, which may provide a reasonable explanation of the decrease in isolation observed in our experiment. 

\begin{table*}
	\caption{Testing results of circulators. All measurements are at $1550~\nano\meter$.}
	\label{tab:CIRC_res}
	\begin{tabular}[t]{@{\extracolsep{1ex}}l@{}c@{}c@{}c@{}c@{}c@{}c@{}c@{}c@{}c@{}c@{}}
		\hline\hline
		& & \multicolumn{4}{c}{Initial} \\
		\cline{3-6}
		\multirow{2}{*}{Sample}  & \makecell{Specified minimum \\ isolation for\\ all ports ($\deci\bel$)} 
		& \multicolumn{2}{c}{\makecell{Insertion\\ loss ($\deci\bel$)}} & \multicolumn{2}{c}{\makecell{Isolation ($\deci\bel$)}}
		& \multicolumn{2}{c}{\makecell{Minimum\\ isolation ($\deci\bel$)}}
		& \multicolumn{2}{c}{\makecell{Maximum\\ decrease of\\ isolation ($\deci\bel$)}}
		& \multirow{2}{*}{\makecell{Irreversible\\ damage at}} \\
		\cline{3-4} \cline{5-6} \cline{7-8} \cline{9-10}
		& & 1 to 2 & 2 to 3 & 2 to 1 & 3 to 2 & 2 to 1 & 3 to 2 & 2 to 1 & 3 to 2 \\
		\hline
		CIR 1			& 45	& 1.03	& 1.07	& 61.4	& 60.6	& 34.7 @ $3.6~\watt$	& 32.2 @ $3.6~\watt$	& 26.7						& 28.4	& was not tested							\\
		CIR 2			& 40	& 0.72	& 0.83	& 67.0	& 65.7	& 38.3 @ $4.6~\watt$	& 32.3 @ $4.6~\watt$	& 28.7						& 33.4	& $4.6~\watt$, $910~\second$	\\
		CIR PM 3	& 25	& 1.00	& 0.80	& 37.0	& 27.0	& was not tested			& 6.4 @ $0.7~\watt$		& was not tested	& 20.6	& $0.9~\watt$, $90~\second$		\\
		\hline\hline
	\end{tabular}
\end{table*}

\subsection{Test results for fiber-optic circulators}
\label{sec:results-circulator}

We have tested three fiber-optic circulators. Samples of CIR~1 and CIR~2 are polarization-insensitive, while CIR~PM~3 is polarization-dependent. Similar to the isolators, the specified operating power is $500~\milli\watt$ for CIR~1 and CIR~2 ($300~\milli\watt$ for CIR~PM~3), and the operating temperature range is from 0 to $+70~\celsius$.

A summary of our testing results is given in~\cref{tab:CIRC_res}. The isolation is temporarily reduced not only between the ports illuminated by the laser (from port~3 to port~2) but also between the unilluminated ports~(from port~2 to port~1). Specifically, the isolation from port~3 to port~2~(port~2 to port~1) decreases by $20.6$--$33.4~\deci\bel$~($26.7$--$28.7~\deci\bel$) at maximum. The residual isolation is $6.4$--$32.3~\deci\bel$ from port 3 to port 2 and $34.7$--$38.3~\deci\bel$ from port~2 to port~1, which is lower than the minimum isolation specified by the component manufacturer for all the samples. Thus, the transmission paths from port 1 to port 2 and from port~2 to port~3 are vulnerable to Eve's high-power injection attack.

\begin{figure}
  \includegraphics{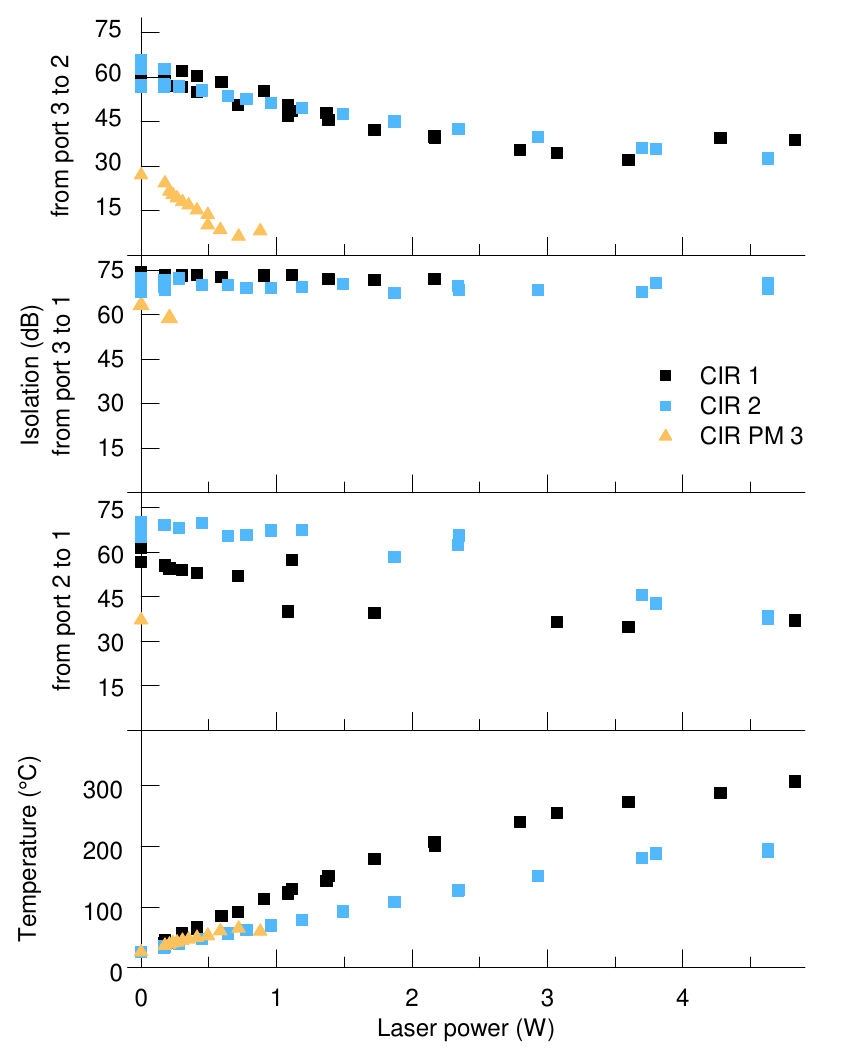}
  \caption{Circulators' values of isolation and the maximum surface temperature under testing. Each point represents the minimum isolation achieved under each applied power.}
  \label{fig:CIRC_data}
\end{figure}

\begin{figure}
	\includegraphics{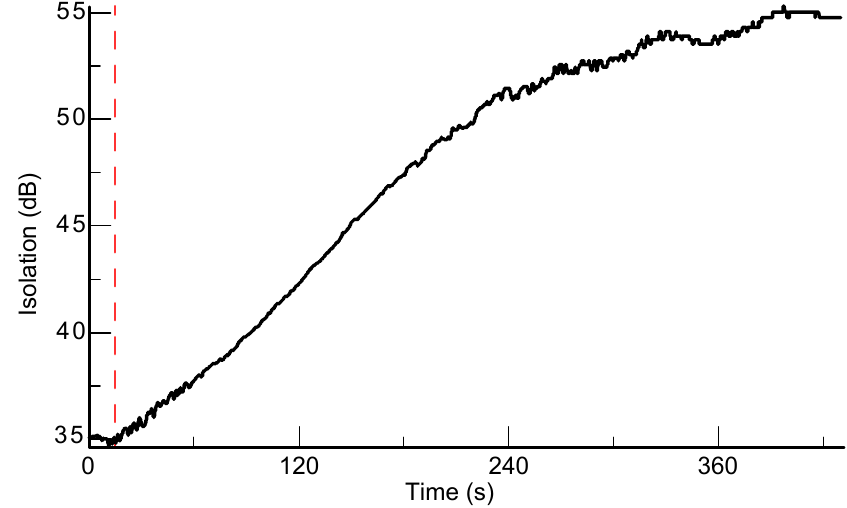}
  \caption{Isolation of CIR~1 from port~2 to port~1 recovers gradually after illumination by $3.6~\watt$ laser power. The vertical dashed line is the HPL's switch-off time.}
  \label{fig:CIRC_trend}
\end{figure}

The detailed measurement data are presented in \cref{fig:CIRC_data}, showing isolation from port~3 to port~2, from port~3 to port~1, and from port~2 to port~1, as well as the maximum surface temperature under different illumination powers. The values of isolation from port~3 to port~2 and port~2 to port~1 are obviously decreased with the increased laser power, as the coupling ratio mainly depends on the polarization rotation provided by a Faraday mirror inside the circulator. However, the isolation from port~3 to port~1 remains essentially unchanged under all experimental conditions for all the tested samples, which is due to no coupling between these two ports according to the internal scheme. Similar to the isolators under test, the temperature of the sample's surface also rises with the laser power.

For both polarization-insensitive samples, the minimum remaining isolation from port~3 to port~2 is $32.2~\deci\bel$ (CIR~1) and $32.3~\deci\bel$ (CIR~2) at the laser power of $3.6$ and $4.6~\watt$ respectively. After that, the isolation value rises for CIR~1, and we thus stop our testing of it at the laser power of $4.8~\watt$ without observing an irreversible damage. Meanwhile, irreversible damage happens for CIR~2 with the increase in its insertion loss to $2.5~\deci\bel$ from port~2 to port~3 at $4.6~\watt$.

Moreover, for each of these two samples, we have measured the isolation from port~2 to port~1 immediately after the laser exposure and found that the sample's heating also temporarily reduces it. Take CIR~1 as an example. \Cref{fig:CIRC_trend} illustrates its recovery after the laser exposure, in which the value of isolation reduces to about $35~\deci\bel$ once after being illuminated by the HPL. After the HPL is switched off, the isolation then recovers to $55~\deci\bel$ in $400~\second$, during which the sample's surface temperature decreases from $272$ to $44~\celsius$.

Surprisingly, for the polarization-sensitive sample, CIR~PM~3, the isolation from port~3 to port~2 falls rapidly with the increased laser power, dropping to only $6.4~\deci\bel$ at the input laser power of $700~\milli\watt$. At $900~\milli\watt$, the insertion loss from port~2 to port~3 increases irreversibly to $15.5~\deci\bel$. Since port~2 and port~3 of this sample is supposed to be used in the QKD system purely as an isolator, we have not measured the change in the insertion loss from port~1 to port~2, isolation from port~2 to port~1, and isolation from port~3 to port~1.

\section{Discussion and countermeasures}
\label{sec:countermeasure}

The experimental results shown above provide two opposite insights into the security of a QKD system. We first discuss the hacking aspect about the vulnerabilities in a QKD system caused by the isolation reduction of the tested isolators and circulators. Then, from the defence point of view, we propose a possible countermeasure to protect the QKD source from these vulnerabilities.

The isolation reduction introduced by high-power laser opens loopholes for at least two possible attacks on QKD, the Trojan-horse attack~\cite{gisin2006,jain2014} and the laser-seeding attack~\cite{huang2019,sun2015,pang2020}. Regarding the Trojan-horse attack, the isolation of the source strongly impacts the secure key rate and transmission distance. The reduced isolation of the source allows Eve to inject more Trojan-horse light into Alice, which is assumed to linearly increases the reflection light. Given $15.2$--$34.5~\deci\bel$ decrease in isolation obtained from our testing results, the photon number of reflection pulse increases by about $2$--$3$ orders from the safe value. These amounts of increase in leaked photon number result in the maximum transmission distance shortens by $20$--$100~\kilo\meter$ according to the various, theoretical security analysis \cite{lucamarini2015,tamaki2016,wang2018,navarrete2022}.

Regarding the laser-seeding attack, an injection power in the order of $100~\nano\watt$ after passing the built-in isolator of Alice's laser to reach the laser cavity is sufficient for achieving a successful attack~\cite{huang2019}. According to our experimental result, the maximum power transmitted through the isolation component is $190~\milli\watt$, assuming the injected power is $10~\watt$~\cite{huang2020} and the isolation is reduced to $17.2~\deci\bel$ as in ISO~PM~2. (Although the minimum value of isolation obtained in our experiment is $6.4~\deci\bel$ for CIR~PM~2, we exclude this type of circulator from the analysis owing to its poor performance and we do not recommend it for use in QKD systems.) To prevent the laser-seeding attack, other components in QKD source should provide about $62.7~\deci\bel$ of isolation. Assuming the built-in isolation of the laser is typically $30~\deci\bel$, the success of the laser-seeding attack relies on the attenuation value of an optical attenuator in Alice. If the attenuation value is less than $32.7~\deci\bel$, the security of the QKD system might be compromised under laser-seeding attack. It is notable that in the above analysis, we assume that the attenuator in Alice works as designed, which does not affect the effectiveness of above-mentioned attacks. Regarding to the possible vulnerability of attenuators, the decreased attenuation under laser-damage attack has been investigated in~Ref.~\onlinecite{huang2020}.

Most importantly, our study also provides a possible countermeasure against the light-injection attacks---adding an extra isolation component into the source unit to be the first one illuminated by the injected light. Its minimum residual isolation upper-bounds the maximum power that can transmit through to reach other optical components. Specifically, the minimum observed isolation is $6.4~\deci\bel$ and $17.2~\deci\bel$ for the polarization-dependent circulator CIR~PM~3 and isolator ISO~PM~2, respectively. Typical minimum residual isolation is more than $20~\deci\bel$ for all the polarization-insensitive components. Therefore, the injected power is limited to less than $190~\milli\watt$, which cannot successfully conduct the laser-damage attack on any optical components according to the previous testing~\cite{bugge2014,makarov2016,huang2020}. If the attacker attempts to further increase the illumination power, the first component fails permanently with a very high insertion loss, which results in a denial of service and thus protects the QKD system from the leakage of secret information~\cite{makarov2016}. Moreover, the isolation required for protection against the Trojan-horse attack and the laser-seeding attack should be calculated starting from the component behind this sacrificial isolator or circulator. Therefore, the extra isolator or circulator placed at Alice's output would protect the rest of the QKD source against the light-injection attacks.

\section{Conclusion}
\label{sec:conclusion}

In this paper, we study the effect of high-power laser on the fiber-optics isolators and circulators and propose the effective countermeasure against light-injection attacks on a QKD system. This study first raises awareness of insecure isolation components---isolators and circulators---in QKD systems. Specifically, the testing shows that the values of isolation provided by the optical isolators and circulators under test are reduced to $17.2$ and $6.4~\deci\bel$ at minimum when high-power laser light is injected into them in the reverse direction. This decrease of isolation opens loopholes, which may allow Eve to conduct the Trojan-horse attack, the laser-seeding attack, and possibly other attacks that inject light into the source. The testing methodology proposed in this study is general and applicable to the other commercial fiber-optic isolators and circulators. To enhance the protection of the QKD source unit, an extra isolation component, an optical isolator or circulator, is needed to defeat the light-injection attacks. The residual isolation of this extra component is sufficient to protect the other components behind it. Any isolation calculated for countermeasure against the Trojan-horse attack and the laser-seeding attack shall be started from the components behind this sacrificial isolation component. Our study shows that the source unit in the QKD system needs this additional layer of protection to be secure.

\acknowledgments
We thank K.~Wei, F.~Xu, and our industry partners for providing us device samples. This work was funded by the Ministry of Science and Education of Russia (programs 5-in-100, NTI center for quantum communications, and grant 075-11-2021-078), Russian Science Foundation (grant 21-42-00040), Canada Foundation for Innovation, MRIS of Ontario, the National Natural Science Foundation of China (grants 61901483 and 62061136011), the National Key Research and Development Program of China (grant 2019QY0702), and the Research Fund Program of State Key Laboratory of High Performance Computing (grant 202001-02). P.C.\ acknowledges support from the DPST scholarship and NSRF via the Program Management Unit for Human Resources \& Institutional Development, Research and Innovation (grant B05F640051).

{\em Author contributions:} A.P.,\ D.R.,\ V.E.,\ A.H.,\ and P.C.\ conducted the experiment. A.P.,\ D.R.,\ V.E.,\ A.H.,\ and V.M.\ analysed the data. A.P.,\ D.R.,\ and A.H.\ wrote the article with help from all authors. A.H.\ and V.M.\ supervised the project.

\appendix

\section{Theoretical temperature dependence of the Verdet constant and isolation in fiber-optic isolators}
\label{sec:theory-isolation-vs-temp}	

\setcounter{figure}{0}
\numberwithin{figure}{section}
\numberwithin{equation}{section}

\subsubsection{Faraday effect in a polarization-dependent isolator}
\label{sec:Faraday}

\begin{figure}
	\includegraphics{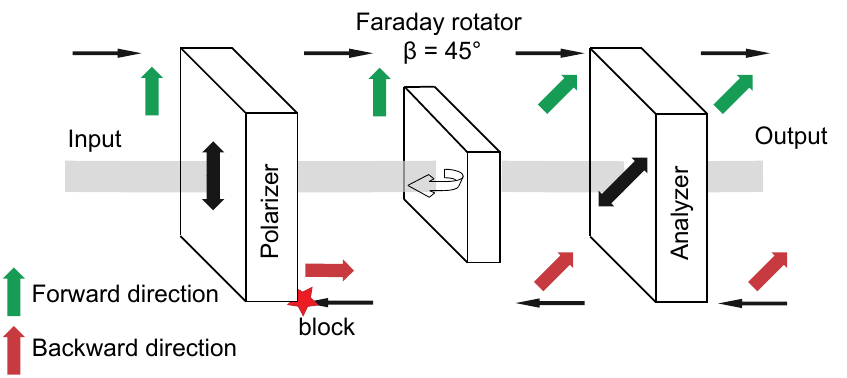}
	\caption{Optical configuration of a polarization-sensitive optical isolator.}
	\label{fig:Faraday-effect}
\end{figure}

An optical isolator is a component that only allows unidirectional transmission of the optical signal. The principal scheme of polarization-dependent isolator is shown in~\cref{fig:Faraday-effect}. It consists of an input polarizer, a Faraday rotator, and an output polarizer called an analyzer. The optical axis of the second polarizer is oriented at an angle $\beta = 45\degree$ with respect to the first polarizer. In this configuration, the optical signal coming from the left side passes through the first polarizer whose optical axis is in the vertical direction, which matches the polarization of the input optical signal. Then a Faraday rotator rotates the polarization of the optical signal by $45\degree$ in a clockwise direction. If there is an introduced laser beam from the optical circuit on the right side, this optical signal has to pass through the Faraday rotator from right to left. Since the Faraday rotator is a non-reciprocal device, the polarization state of the reflected optical signal will rotate for an additional $45\degree$ in the same direction as the input signal, thus becoming perpendicular to the optical axis of the first polarizer. 

As shown above, an optical isolator is based on the Faraday effect~\cite{zvezdin1997}. The polarization plane of linearly polarized light beam during propagation in a magneto-optical crystal is rotated by an angle $\theta$. The direction of rotation is dependent on the direction of the magnetic field and not on the direction of light propagation. The relation between the angle of polarization rotation and the magnetic field in a crystal is
\begin{equation}\label{eq:polarization rotation}
\theta = V(\lambda, T)BL,
\end{equation}
where $B$ is the longitudinal magnetic field component in$~\tesla$, $L$ is the length of the path where the light and magnetic field interact in$~\meter$, and $V(\lambda, T)$ is the Verdet constant depending on the wavelength of the propagating light $\lambda$ and temperature of magneto-optic crystal $T$ in $\rad$/$(\tesla \cdot \meter)$. Here we shall consider only the temperature dependence.  

The temperature dependence of Verdet constant and hence the angle of Faraday rotation leads to variation of the isolation coefficient with temperature. Modern single-mode isolators have a high stability of isolation in the temperature range from~$5$ to~$70~\celsius$. Thermal effects can be neglected for typical optical circuits, such as QKD systems, with laser power less than $300$--$500~\milli\watt$. However, when the high-power laser is applied in the reverse direction, its emission is partially absorbed inside the isolator and induces heating of the magneto-optic crystal~\cite{kiriyma2015}. The temperature dependence of the Verdet constant causes the changes of the angle of polarization plane rotation [see~\cref{eq:polarization rotation}] \cite{snetkov2014,khazanov2016}. For optical isolators, it means reducing the isolation coefficient in the reverse direction and losing power and degraded beam quality in the forward direction. Thermal effects can be mitigated by a careful choice of the magneto-optical material in the component~\cite{snetkov2014}. The most widespread materials for a single-stage fiber isolator in near infrared band are rare-earth garnets~\cite{booth1984,mukimovb1990,khazanov2016}. Here we consider the following types of garnets: yttrium iron garnet (YIG), terbium gallium garnet (TGG), and bismuth-substituted yttrium iron garnet (Bi:YIG).
\subsubsection{Verdet constant model}
\label{sec:Verdet model}

In a general case, the Verdet constant of rare-earth garnet is impacted by several different contributions~\cite{slezak2016,serber1932,buckingham1966}. In our case, only temperature-dependent contributions are considered: the paramagnetic contribution $V_{pm}$~(for more detail see Ref.~\onlinecite{vojna2018}) and frequency-independent gyromagnetic term $V_{gm}$ (detail in Ref.~\onlinecite{zvezdin1997}). The Verdet constant as a function of temperature has the appearance
\begin{equation}\label{eq:Verdet_f}
V(T)= V_{pm}+V_{gm} = -\dfrac{A\lambda_{0}^2}{(T-T_{w})}+\dfrac{B}{T-T_{w}}+ C,
\end{equation}
where $\lambda_{0}$ is the wavelength of dominant electronic transition, $T_{w}$ is the Curie temperature, and $A$, $B$, $C$ are constants depending on the properties of chosen material. 

\begin{figure}
	\includegraphics{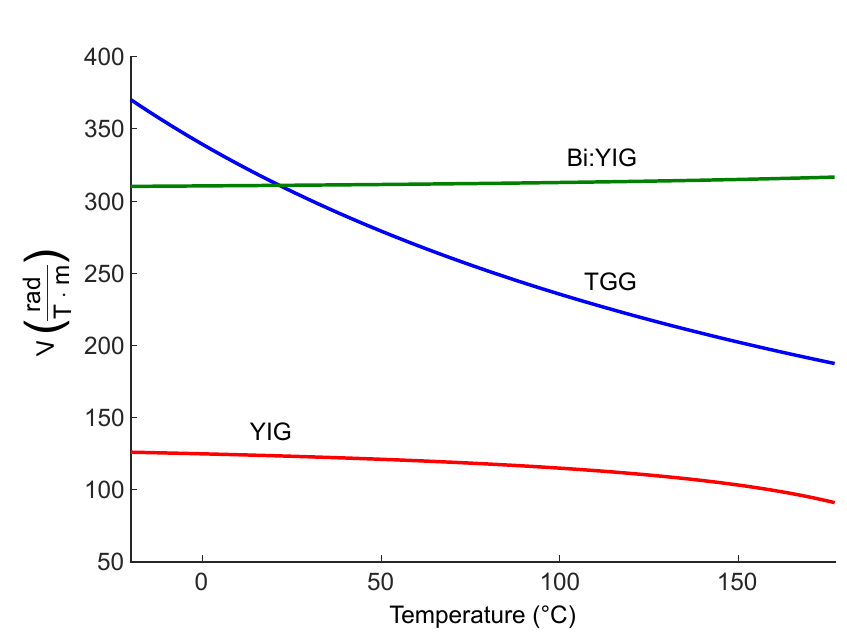}
	\caption{The temperature dependence of the Verdet constant for TGG, YIG, and Bi:YIG, at $1550~\nano\meter$ wavelength.}
	\label{fig:Temperature dependence of the Verdet}
\end{figure}

Using data from Refs.~\onlinecite{vojna2018,zao2019,cooper1968,crossley1969,stevens2016,matsumoto1986,vertruyen2008,vojna2019,kumari2018}, the dependence of the Verdet constant is obtained within a temperature range from $-20$ to $175~\celsius$, as presented in~\cref{fig:Temperature dependence of the Verdet}. These dependencies have been calculated with fixed operating wavelength $\lambda = 1550~\nano\meter$. As reflected in~\cref{fig:Temperature dependence of the Verdet}, the crystal TGG has exhibited the least stability with temperature. This means that isolators based on TGG are most susceptible to thermal effects at $\lambda = 1550~\nano\meter$. Isolators based on YIG or Bi:YIG should be more temperature-stable.

\subsubsection{Isolation model}
\label{sec:Imodel}

\begin{figure}
	\includegraphics{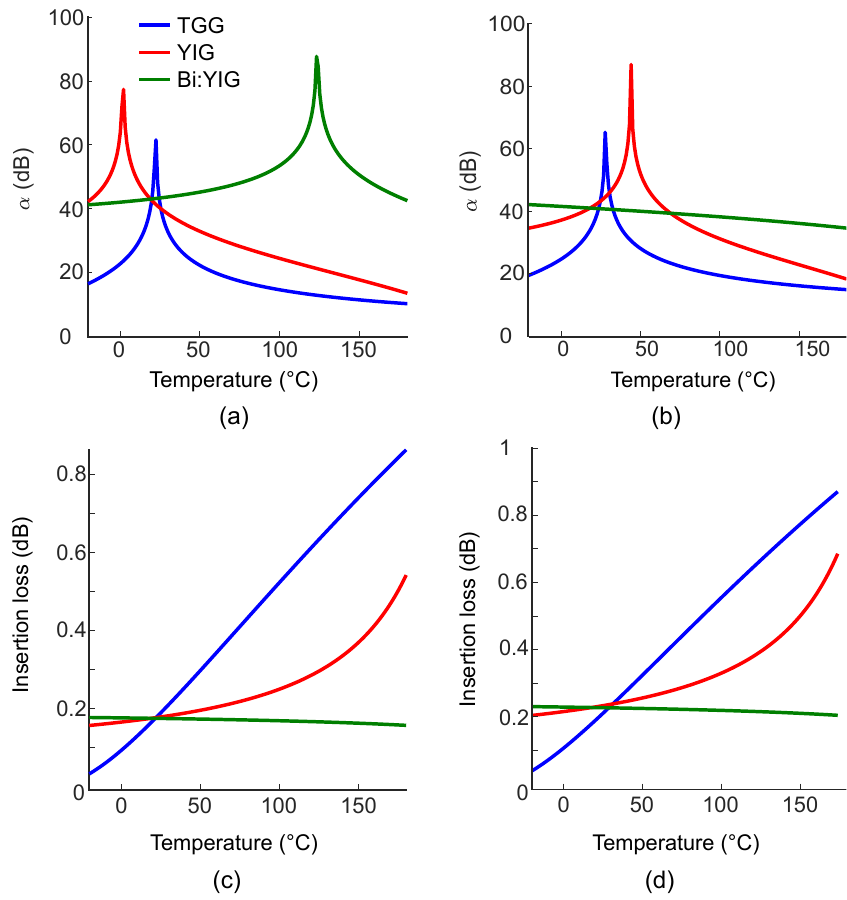}
	\caption{Dependence of the isolation coefficient (a) and (b) and insertion loss (c) and (d) on temperature for TGG, YIG, and Bi:YIG. (a) and (c) correspond to $\theta = 44.43\degree$; (b) and (d) correspond to $\theta = 45.57\degree$.}
	\label{fig:Dependence of isolation}
\end{figure}

Next, we analyze the change in isolation with varying crystal temperature in the proposed model with the ideal polarizer and analyzer. The polarization planes of the polarizer and the analyzer are oriented relative to each other at the angle~$\beta$, and Faraday rotator provides the $45\degree$ rotation of the polarization plane of the propagating light with a central wavelength of $1550~\nano\meter$. In our model, the magnetic field is constant and independent of temperature (but in real systems magnetic field might introduce changes in isolation). According to Malus's law, after passing through Faraday rotator and the polarizer, the intensity of a beam of plane-polarized light varies as $I = I_{0}\cos^2(\theta + \beta)$, where $I_{0}$ is the initial intensity~\cite{booth1984,vojna2018}. The isolation coefficient is then defined as $\alpha = -10 \log \cos^2 (\beta + \theta)$. (The insertion loss may be found from the similar formula using rotation angles equal to $(\beta-\theta)$.) After substituting the value of $\theta$ from~\cref{eq:polarization rotation}, the temperature dependence of the isolation coefficient takes the form
\begin{equation}\label{eq:Isolation}
\alpha(T)=-10 \log \left[ \beta + \dfrac{V(T)}{V(25~\celsius)} \cdot k \right],
\end{equation}
where $k$ is the coefficient depending on the initial isolation value at the temperature of~$25~\celsius$~\cite{konno1993}. Let's use the initial isolation value of $40~\deci\bel$, as a typical value for single-stage isolators at room temperature ranges from $32$ to $40~\deci\bel$ according to their specification~\cite{konno1993}. The isolation of~$40~\deci\bel$ corresponds to the rotation angle of polarization plane in the Faraday rotator either $\theta = 44.43\degree$ or $\theta = 45.57\degree$, depending on the direction of rotation. The calculation results for isolation and insertion loss are presented in~\cref{fig:Dependence of isolation}.

The model predicts sharp peaks in the isolation value. It should be noted that generally there are no pronounced peaks in our experimental results of the isolation coefficient when the components being heated by the laser. This may be explained by the internal scattering in the crystal, which leads to a partial change in the plane of polarization. However, this factor is not considered in this model. 

\subsubsection{Outcome}
\label{sec:Out}

\begin{figure}
	\includegraphics{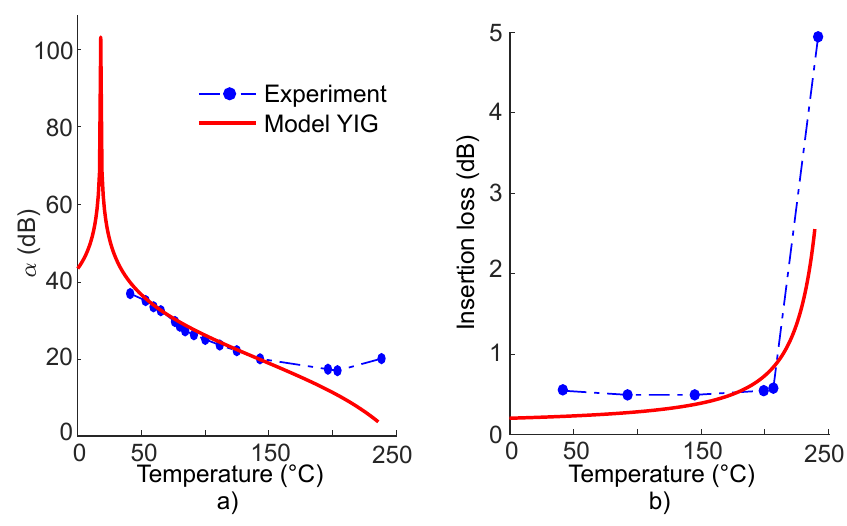}
	\caption{Comparison of experimental results for ISO~PM~2 and model for YIG with $\theta = 44.43\degree$, for (a) isolation coefficient and (b) insertion loss.}
	\label{fig:Compare results}
\end{figure}

Our model shows that the change of the isolation coefficient with temperature depends heavily on the material of the magneto-optical crystal, even though each garnet may provide the same isolation value at room temperature. The crystal TGG has demonstrated the sharpest decrease in the isolation coefficient in the operating temperature range of isolators. This is because the operating wavelength range for this garnet is from~$700$ to~$1100~\nano\meter$~\cite{vojna2018}. The Bi:YIG crystal is specially designed for applications demanding high values of the isolation coefficient over a wide temperature range~\cite{kiriyma2015,vojna2019}. According to the calculation, the isolation coefficient is more than~$40~\deci\bel$ in the temperature range from $-20$ to $180~\celsius$. Such a high isolator stability is achieved due to the optimal crystal composition~\cite{zao2019}. Additional doping provides several sublattices in the crystal structure, which compensate the temperature dependence of the Verdet constant (and isolation respectively) of each other. In YIG, our model predicts that the isolation decreases by about~$10~\deci\bel$ at $70~\celsius$. When temperature increases significantly (up to $175~\celsius$), isolation drops to about $15~\deci\bel$. The obtained result fits well with the experimental data for ISO~PM~2. The comparison of experiment with model is shown in~\cref{fig:Compare results}.

In summary, our model shows that Bi:YIG has the weakest dependence of the isolation coefficient on temperature and therefore it is the most advanced garnet for the isolators resilient to the laser-damage attack. In addition, we may assume that the magneto-optic crystal in the isolator ISO~PM~2 is YIG.

\def\bibsection{\medskip\begin{center}\rule{0.5\columnwidth}{.8pt}\end{center}\medskip} 

\end{document}